\documentclass[a4paper,11pt,aps,tightenlines,nofootinbib]{revtex4}
\usepackage{xspace}
\usepackage[english]{babel}
\usepackage{amsfonts}
\usepackage{amsmath, amsthm, amssymb, mathrsfs}
\usepackage[dvips]{graphicx}
\usepackage{pdfpages}
\usepackage{setspace}
\usepackage{indentfirst}
\usepackage[margin=2.5cm]{geometry}

\usepackage{pstricks}
\usepackage{enumerate}
\usepackage{pstricks-add}
\usepackage{epsfig}
\usepackage{pst-grad}
\usepackage{pst-plot}

\newcommand{\abs}[1]{\left|#1\right|}
\newcommand{\bra}[1]{\left< #1 \right|}

\newcommand{\ket}[1]{|#1\rangle}
\newcommand{\norm}[1]{\left|#1\right|}

\newcommand{\refeq}[1]{eq.(\ref{#1})}

\newcommand{\be}{\begin{equation}}
\newcommand{\ee}{\end{equation}}
\newcommand{\ba}{\begin{array}{c}}
\newcommand{\ea}{\end{array}}

\begin{document}

\title{A curvature operator for LQG}
\author{E. Alesci}
\email[]{emanuele.alesci@fuw.edu.pl}
\author{M. Assanioussi}
\email[]{mehdi.assanioussi@fuw.edu.pl}
\author{J. Lewandowski}
\email[]{jerzy.lewandowski@fuw.edu.pl}
\affiliation{Institute of Theoretical Physics, University of Warsaw
(Instytut Fizyki Teoretycznej, Uniwersytet Warszawski), ul. Ho{\.z}a 69, 00-681 Warszawa, Poland, EU}

\begin{abstract}
We introduce a new operator in Loop Quantum Gravity - the $3D$ curvature operator - related to the $3$-dimensional scalar curvature. The construction is based on Regge Calculus. We define this operator starting from the classical expression of the Regge curvature, we derive its properties and discuss some explicit checks of the semi-classical limit.
\end{abstract}

\maketitle

\section{Introduction}

Loop Quantum Gravity \cite{lqgcan1} is a promising candidate to finally realize a quantum description of General Relativity. The theory presents two complementary descriptions based on the canonical and the covariant approach (spinfoams) \cite{lqgcov}. The first implements the Dirac quantization procedure \cite{DiracQM1964} for GR in Ashtekar-Barbero variables \cite{AB variables} formulated in terms of the so called holonomy-flux algebra \cite{lqgcan1}: one considers smooth manifolds and defines a system of paths and dual surfaces over which the connection and the electric field can be smeared. The quantization of the system leads to the full Hilbert space obtained as the projective limit of the Hilbert space defined on a single graph.
The second is instead based on the Plebanski formulation \cite{Plebanski:1977zz} of GR, implemented starting from a simplicial decomposition of the manifold, i.e. restricting to piecewise linear flat geometries. Even if the starting point is different (smooth geometry in the first case, piecewise linear in the second) the two formulations share the same kinematics \cite{Kaminski:2009fm} namely the spin-network basis \cite{Rovelli:1995ac} first introduced by Penrose \cite{Penrose}.
In the spinfoam setting, with its piecewise linear nature, a beautiful interpretation of the spin-networks in terms of quantum polyhedra \cite{cov volume1} naturally arises. This interpretation is not needed in the canonical formalism where continuous geometries lead to polymeric quantum geometries. However in \cite{Freidel Geiller} it has been proven that the discrete classical phase space (on a fixed graph) of the canonical approach based on the holonomy-flux algebra can be related to the symplectic reduction of the continuous phase space respect to a flatness constraint; this construction allows a reconciliation between the loop gravity geometrical interpretation in terms of singular geometry, and the spin foam interpretation in terms of piecewise flat geometry, since it can be shown that both geometries belong to the same equivalence class.
Canonical LQG and Spinfoam appear much closer if we allow in the first to disentangle the discretization from the quantization procedure.
In this article we want to pursue this perspective as a tool to build a curvature operator in LQG, fundamental to solve the most challenging issue in the canonical approach: the quantum dynamics related to the Hamiltonian constraint.
The Hamiltonian Constraint has been quantized by Thiemann \cite{Thiemann:1996ay, Thiemann96a} improving several previous proposals \cite{Early-hamiltonians} and finally succeeding in defining an anomaly free operator. It is defined employing a regularization procedure with specific rules that might be changed to bring it closer to the spinfoam formalism \cite{iopro, ioreg} (but till now spoiling the anomaly free property). However this operator is computationally extremely hard to implement \cite{Alesci:2013kpa}, in particular its Lorentzian part which involves several commutators of the extrinsic curvature in order to express the Ricci scalar in terms of Holonomies and Fluxes. Few computations appeared so far \cite{Borissovetal97, Gaul} and few solutions have been found \cite{ioantonia}.
The idea developed in this paper is the following: the Lorentzian term of the Hamiltonian constraint, that is just the integral of the Ricci scalar over the $3$-dimensional surface of the foliation $\int_{\Sigma}{\sqrt q R}$, can be seen as the Einstein-Hilbert (E-H) action in $3D$ and we know how to write this expression using Regge Calculus \cite{Regge1,Regge2} in terms of geometrical quantities, i.e. lengths and angles. Operators of length \cite{lunghezza1, lunghezza2, lunghezza3} and angle \cite{angoli1, angoli2, angoli3} are available in LQG and Spinfoams and by using a suitable regularization procedure for the classical quantities we can implement directly the integral of the Ricci scalar as an operator acting on spin-network states and in this way settle the first step to bypass many of the complications appearing in the Lorentzian constraint. The considerations due to the change of regularization procedure bring us close to the perspective of \cite{Freidel Geiller} and make our proposal intriguing 
also for 
the spinfoam formalism.\\
The article is organized as follows:
In the first section we present briefly Regge calculus, discuss the generalization of the $3D$ Regge action to arbitrary piecewise flat decomposition and we highlight some results about the convergence of the discrete action to the continuum one. In the second section we present our construction of the $3D$ scalar curvature operator writing the classical formulas for the length and the angle in terms of fluxes and we expose the adopted regularization scheme. Those geometrical quantities are then promoted to operators and by means of an averaging procedure we build the final expression of the gauge invariant curvature operator. Finally, in the third section we discuss some properties of this operator and its semi-classical (large spins limit) behavior.

\section{Scalar curvature for a piecewise flat space}

\subsection{Regge calculus}\label{sec11}

Regge calculus \cite{Regge1,7} is a discrete approximation of general relativity which approximates spaces with smooth curvature by piecewise flat spaces: given a $n$-dimensional Riemannian manifold $\Sigma$, we consider a simplicial decomposition $\Delta$ ``approximating'' $\Sigma$ where we assume that curvature lies only on the \emph{hinges} of $\Delta$, namely on its $n-2$ simplices. In this context, Regge derived the simplicial equivalent of the E-H action:
\begin{equation} \label{e1}
S_{EH}=\int_\Sigma \sqrt{-g} R \;d^nx \quad \rightarrow \quad S_R=\int_\Delta \sqrt{-g} R \;d^nx =2\sum_h \epsilon_h V_h
\end{equation}
where the sum extends to all the hinges $h$ with measure $V_h$ and deficit angle $\epsilon_h$:
\begin{align}\label{e2}
 \epsilon_h &= 2\pi -\sum_{s_h} \theta_h^{s_h}=\sum_{s_h}\left(\frac{2\pi}{\alpha_h}-\theta_h^{s_h} \right) & \qquad &\text{if the hinge $h$ is not on the boundary}& \\
 \nonumber \epsilon_h &= \pi  -\sum_{s_h} \theta_h^{s_h}=\sum_{s_h}\left(\frac{2\pi}{\alpha_h}-\theta_h^{s_h} \right) & \qquad  &\text{if the hinge $h$ is on the boundary}&
\end{align}
$\theta_h^{s_h}$ is the dihedral angle at the hinge $h$ and the sum extends to all the simplices $s_h$ sharing the hinge $h$. The coefficient $\alpha_h$ is the number of simplices sharing the hinge $h$ or twice this number if the hinge is respectively in the bulk, or on the boundary of the triangulation.
Using simplices for the decomposition implies that both $V_h(l_{ab})$ and $\epsilon_h(l_{ab})$ are functions of the hinges lengths $l_{ab}$ joining two sites $a$ and $b$ of $\Delta$.
Equation \eqref{e1} can also be written in another form which, as we will see later, is more adapted to our quantization scheme:
\begin{equation} \label{e3}
\frac{1}{2}\int_\Delta \sqrt{-g} R \;d^nx =\sum_h \sum_{s_h} V_h^{s_h} \left(\frac{2\pi}{\alpha_h}-\theta_h^{s_h} \right)=\sum_s \sum_{h \in s} V_h^{s} \left(\frac{2\pi}{\alpha_h}-\theta_h^{s} \right)
\end{equation}
Where in the last equality the first sum is over simplices $s$ of $\Delta$ while the second is over the hinges $h$ in each simplex.

The purpose of this work is to define a scalar curvature operator for LQG implementing a regularization of $S_{EH}$ in terms of a simplicial decomposition that allows to replace $S_{EH}$with the right-hand side of \eqref{e3} and finally promote this expression to a well defined operator acting on the LQG kinematical Hilbert space. As we are interested in the Hamiltonian constraint of the $4$-dimensional theory, we consider only spaces of dimension $n=3$. Therefore the expression we want to quantize is:
\begin{equation}\label{e4}
 \frac{1}{2}\int_\Delta \sqrt{-g} R \;d^3x =\sum_s \sum_{h \in s} L_h^{s} \left(\frac{2\pi}{\alpha_h}-\theta_h^{s} \right)
\end{equation}
where $L_h^{s}$ is the length of the hinge $h$ belonging to the simplex $s$.

Also, if we think only about computing the integral of the scalar curvature, the formula \eqref{e4} can be extended to arbitrary piecewise flat cellular decompositions as presented below. This is an important step in our approach to construct the operator as the reason will be clear later.
\\

\subsection{From simplicial decompositions to arbitrary piecewise flat cellular decompositions}\label{sec12}

This section is about generalizing the classical Regge expression for the integrated scalar curvature. It is important to point out that we are only interested in computing the quantity $\int \sqrt{\abs{g}} R\ dx^3$ for an arbitrary piecewise flat cellular decomposition of space. This does not mean that we are building a generalization of Regge calculus as we don't derive any equations of motion.\\

Let us first introduce the definition of a cellular decomposition: a cellular decomposition $\mathscr{C}$ of a space $\Sigma$ is a disjoint union (partition) of open cells of varying dimension satisfying the following conditions:\\
{\it i) An n-dimensional open cell is a topological space which is homeomorphic to the n-dimensional open ball.\\
ii) The boundary of the closure of an n-dimensional cell is contained in a finite union of cells of lower dimension.\\}

In $3D$ Regge calculus we consider a simplicial decomposition of a $3D$ manifold which is a special cellular decomposition. Using the $\epsilon$-cone structure \cite{Regge1} we induce a flat manifold with localized conical defects. Those conical defects lie only on the 1-simplices and encode curvature. Thereby it can be proven that scalar curvature is distributional and proportional to the deficit angles carried by the 1-simplices. Then by integration over the entire space one gets equation \eqref{e1} (see \cite{7}). This construction is independent from the choice of the simplicial decomposition: the same expression would hold for arbitrary piecewise flat cellular decompositions i.e decompositions such that the space inside each 3-cell is flat. The difference being that the deficit angle along one hinge, though still constant, is not determined by the hinges lengths $l_{ab}$. On arbitrary piecewise flat decompositions the lengths do not form a complete set of variables for the theory and more parameters 
such as the angles are needed. The final expression of the integrated scalar curvature in the general case can be written as

\begin{equation}\label{e6}
\frac{1}{2}\int_\mathscr{C} \sqrt{-g} R \;d^3x=\sum_{h \in \mathscr{C}} L_h \epsilon_h =\sum_{c \in \mathscr{C}} \sum_{i \in c} L_h^c \left(\frac{2\pi}{\alpha_h}-\theta_h^c \right)
\end{equation}
where the first sum now is over the 3-cells $c$ and $\alpha_h$ is the number of 3-cells sharing the hinge $h$ (if it's not on the boundary of $\mathscr{C}$).

\eqref{e6} is the classical formula that we adopt to express the integrated scalar curvature and it is the basis of our construction to define a curvature operator.
\\

\subsection{On the convergence of Regge action}

The question of convergence of Regge action to the E-H action and the relationship between the discrete scheme and the corresponding continuum theory is of crucial importance. There have been extensive studies on this aspect of Regge calculus. In particular, it's possible to derive the Regge action from the E-H one \cite{7} and it was shown \cite{13} that given any lattice, regular or not, the deviation of Regge action from its continuum limit can be expressed as a power series in $l^2$, where $l$ is the typical length of the lattice. This proves that Regge action approaches the E-H action when the typical length goes to $0$
\begin{equation}\label{e7}
 \lim_{l \rightarrow 0} S_R = S_{EH}
\end{equation}
provided that certain general boundary conditions are satisfied. Moreover this convergence result can be generalized to some non-simplicial decompositions. For example if we consider a polyhedral decomposition (or any decomposition with flat hinges), the result is recovered by invoking the simple argument that such decomposition can always be refined using simplices and therefore inducing a simplicial decomposition where the additional hinges carry null deficit angles. For more general decompositions where for instance the hinges are not straight lines, the result is not straightforward. However there exists at least a class of such decompositions for which the convergence holds. A simple example is to consider a decomposition where the hinges are arcs of circles such that the length $\breve{l}_{ab}$ in $\mathbb{R}^n$ of each arc, connecting two sites $a$ and $b$ of the lattice, is proportional to the Euclidean distance $l_{ab}$ in $\mathbb{R}^n$ between the same sites
\begin{equation}\label{e26}
 \breve{l}_{ab}=\xi . l_{ab}
\end{equation}
With a proportionality constant $\xi$ that is the same for all hinges.
Let $\Xi$ be the decomposition with arcs as hinges and characterized by the constant $\xi$, and $\Delta$ the decomposition with straight lines connecting the same sites. Note that in $\Xi$, two sites can be connected by any number $h_{ab}$ of hinges\footnote{The number of hinges could be infinite but we exclude this case. Latter on we will see that the prescription we are considering in the regularization implies that the set of hinges linking two sites forms a 3-cell when the number of those hinges exceeds one. Hence an infinite number of hinges would form a 3-cell with infinite number of faces which in the dual picture would correspond to intertwiners of infinite number of spins.} with equal lengths $\breve{l}_{ab}^k$. Thereby we can generate the set of deficit angles $\breve{\epsilon}_{ab}^k$ for $\Xi$ using the deficit angles ${\epsilon}_{ab}^k$ of $\Delta$ such that for every two connected sites $a$ and $b$ we have \footnote{Such choice is always possible since in the general case the deficit angles are 
not determined by the lengths as it was in the simplicial case. For instance we can take $\breve{\epsilon}_{ab}^k = \frac{\epsilon_{ab}}{\xi . h_{ab}}$}
\begin{equation}\label{e27}
\sum \limits_{k=1}^{k=h_{ab}} \breve{l}_{ab}^k .\breve{\epsilon}_{ab}^k= l_{ab} .\epsilon_{ab}
\end{equation}
Where $k$ labels the different arcs connecting the two sites $a$ and $b$. Hence we can write
\begin{equation}\label{e28}
S_R(\Xi)=2\sum_{h \in \Xi} \breve{l}_h .\breve{\epsilon}_h=2\sum_{h \in \Delta} l_h .\epsilon_h = S_R(\Delta)
\end{equation}
where the index $h$ labels the hinges of the decomposition.

This shows that, from any polyhedral decomposition of space, we can construct an equivalent piecewise flat decomposition characterized by a positive number $\xi$ (larger than $1$) where straight hinges are replaced by arcs. Then, by keeping the coefficient $\xi$ constant in the refinement process, the convergence result can be recovered in this particular non-simplicial case. This example suggests that Regge action written for a non-simplicial decomposition, and specially with non straight hinges, could converge to the continuum action. Since the convergence of the expression \eqref{e6} is crucial for the construction and the interpretation of the operator introduced in this work, we have, a priori\footnote{As long as we don't give a general proof of the convergence result for general decompositions, we may expect that this result doesn't hold for all decompositions. But for our construction, specifically in the regularization scheme, it's enough it exists one class of decompositions with non 
straight hinges which verify the convergence.}, to restrict ourselves only to cellular decompositions allowing this convergence result. Therefore in the rest of the paper the term ``cellular decomposition'' will refer to a piecewise flat cellular decomposition for which Regge action converges to the E-H action.

\section{Construction of the curvature operator}\label{sec2}

We start by writing the classical expressions for the length and the dihedral angle in terms of the densitized triad (electric field).

Given a curve $\gamma$ embedded in a 3-manifold $\Sigma$:
\begin{align}
 \nonumber \gamma \ : [0,1] &\rightarrow \Sigma \\
 \nonumber s &\rightarrow \gamma^a (s)
\end{align}
the length $L(\gamma)$ of the curve in terms of the electric field $E_i$ is:
\begin{equation}\label{e8}
L(\gamma)=\int \limits_0 ^1 ds \sqrt{\delta_{ij} G^i (s) G^j (s)}
\end{equation}
where
\begin{equation}\label{e9}
G^i (s)= \frac{\frac{1}{2}\epsilon^{ijk}\epsilon_{abc}E_j^b E_k^c \dot{\gamma}^a(s)}{\sqrt{\frac{1}{3!}\epsilon^{ijk}\epsilon_{abc}E_i^a E_j^b E_k^c}}
\end{equation}
In \eqref{e9} the $E_i$'s are evaluated at $x^a = \gamma^a (s)$ and $\dot{\gamma}^a (s) = \frac{d\gamma^a (s)}{ds}$.\\

To define the dihedral angle, we consider two surfaces $S^1$ and $S^2$ intersecting in the curve $\gamma$. The dihedral angle between those two surfaces is then:
\begin{equation} \label{e10}
 \theta^{12}=\int \limits_0 ^1 ds\left( \pi - \arccos \left[ \frac{\delta^{jk} E_j^b n_b(S^1,s) E_k^c n_c(S^2,s)}{\norm{E_j^b n_b(S^1,s)}\norm{E_k^c n_c(S^2,s)}} \right] \right)
\end{equation}
where $\norm{E_j^b n_b(S^k,s)}=\sqrt{\delta_{ij} E_i^b n_b(S^k,s) E_j^c n_c(S^k,s)}$ and $n_b(S^k,s)$ is the normal one form on the surface $S_k$\footnote{The normals are always taken to be inwards.}.\\

We can therefore express Regge action in terms of the densitized triad as follows:
\begin{align}\label{e29}
 \nonumber \frac{1}{2}\int_\mathscr{C} \sqrt{-g} R \;d^3x=&\sum_{c \in \mathscr{C}} \sum_{h \in c} L_h^c \left(\frac{2\pi}{\alpha_h}-\theta_h^c \right)\\=&\sum_{c \in \mathscr{C}}\ \sum_{\gamma(s)=:h \in c}\ \int \limits_0 ^1 ds \sqrt{\delta_{ii'} \frac{\frac{1}{2}\epsilon^{ijk}\epsilon_{abc}E_j^b E_k^c \dot{\gamma}^a(s)}{\sqrt{\frac{1}{3!}\epsilon^{ijk}\epsilon_{abc}E_i^a E_j^b E_k^c}} \frac{\frac{1}{2}\epsilon^{i'j'k'}\epsilon_{a'b'c'}E_{j'}^{b'} E_{k'}^{c'} \dot{\gamma}^{a'}(s)}{\sqrt{\frac{1}{3!}\epsilon^{i'j'k'}\epsilon_{a'b'c'}E_{i'}^{a'} E_{j'}^{b'} E_{k'}^{c'}}}} \\ \nonumber & .\int \limits_0 ^1 ds \left(\frac{2\pi}{\alpha_h}-\pi + \arccos \left[ \frac{\delta^{jk} E_j^b n_b(S^1,s) E_k^c n_c(S^2,s)}{\norm{E_j^b n_b(S^1,s)}\norm{E_k^c n_c(S^2,s)}} \right] \right)
\end{align}

The next step is to match Regge calculus context with LQG framework. This is achieved by invoking the duality between spin-networks and quanta of space, that allows to describe for example spin-networks in terms of quantum polyhedra \cite{9,10,	cov volume1}. The second step is to define a regularization scheme for the classical expressions that we have. Those steps are detailed in the following sections.

\subsection{Spin networks and decomposition of space}\label{sec21}
\begin{figure}[ht]
\centering
\includegraphics[width=0.9\textwidth]{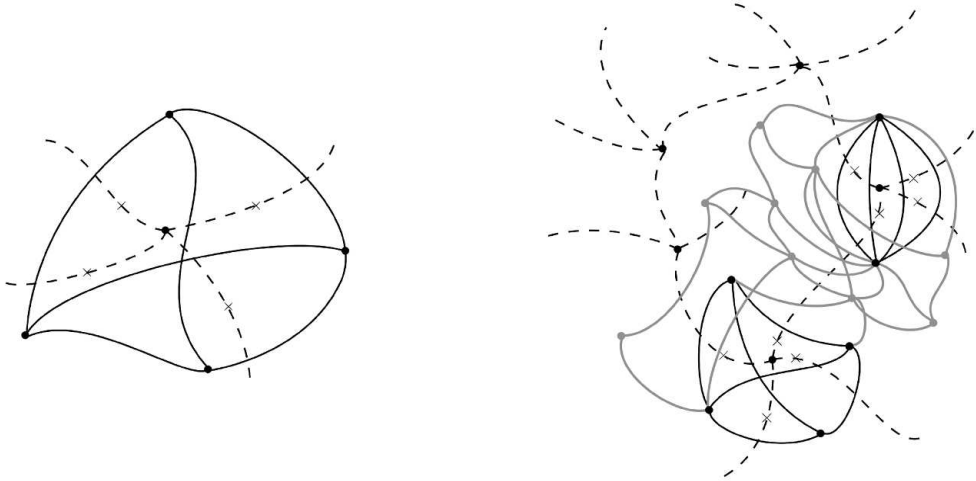}
\caption{\small Examples of covering cellular decompositions: On the left, a 3-cell containing a 4-valent node of $\Gamma$ (dashed line). On the right, a part of a covering cellular decomposition $\mathscr{C}$ around two 4-valent nodes of $\Gamma$ (dashed line): in continuous black lines the 3-cells containing the nodes, in gray the intermediate structure.}
\label{fig1}
\end{figure}

In LQG, we define the kinematical Hilbert space $\mathscr{H}$ of quantum states \cite{lqgcan1} as the completion of the linear space of cylindrical functions $\Psi(\Gamma)$ on all possible graphs $\Gamma$. An orthonormal basis in $\mathscr{H}$ can be introduced, called the spin-network basis, so that for each graph $\Gamma$ we can define a proper subspace $\mathscr{H}_\Gamma$ of $\mathscr{H}$ spanned by the spin-network states defined on $\Gamma$. Those proper subspaces $\mathscr{H}_\Gamma$ are orthogonal to each other and they allow to decompose $\mathscr{H}$ as:
\begin{equation}
 \mathscr{H}= \bigoplus \limits_\Gamma \mathscr{H}_\Gamma
\end{equation}
A spin-network state is defined as an embedded colored graph denoted $\ket{\Gamma,\jmath_l,\imath_n}$, where $\Gamma$ is the graph while the labels $\jmath_l$ are quantum numbers standing for $SU(2)$ representations (i.e spins) associated to edges, and $\imath_n$ are quantum numbers standing for $SU(2)$ intertwiners associated to nodes. It was shown \cite{9,10} that an intertwiner can be seen as the dual to a region of 3D-space with a topologically spherical boundary. This boundary is punctured by the $N$ legs of the intertwiner which means that the boundary surface is made of $N$ elementary patches whose areas is determined by the spins carried by the intertwiner legs (example of 4-valent node on figure \ref{fig1}).

The idea is to use this duality to build a 3-cellular decomposition on a given spin-network graph. Given a spin-network we build a dual cellular decomposition and we will use this to regularize the classical expression \eqref{e29} for the curvature. In the following we give a general prescription to get a cellular decomposition from a spin-network state based only on the spin-network graph $\Gamma$ (different prescription considering also the quantum labels in order to use Minkowski theorem \cite{minko} will be explored in future work). This prescription does not guarantee the uniqueness of the decomposition associated to the graph $\Gamma$, but we don't focus on this issue for now as the rising ambiguities are discussed in details in section \ref{sec25}.

For each spin-network graph we define a {\it covering cellular decomposition} (see figure \ref{fig1}) as follows.

{\it
A cellular decomposition $\mathscr{C}$ of a three-dimensional space $\Sigma$ built on a graph $\Gamma$ is said to be a covering cellular decomposition of $\Gamma$ if:
\begin {enumerate}[i)]
\item  Each 3-cell of $\mathscr{C}$ contains at most one vertex of $\Gamma$;
\item  Each 2-cell (face) of $\mathscr{C}$ is punctured at most by one edge of $\Gamma$ and the intersection belongs to the interior of the edge;
\item  Two 3-cells of $\mathscr{C}$ are glued such that the identified 2-cells match.
\end{enumerate}
}

We can add an additional requirement on the boundary of a 3-cell in order to respect the dual picture introduced to construct the length operator that we need (section \ref{sec22}), this requirement states:
{\it 
\begin {description}
\item [\it iv)] If two 2-cells on the boundary of a 3-cell intersect, then their intersection is a connected $1$-cell.
\end {description}
}

This full set of requirements is quite easy to meet. Furthermore, once a decomposition satisfying these conditions is achieved, subsequent refinements needed in the limiting procedure always exist. Nonetheless, these conditions do restrict the allowed decompositions. As we will see, they ensure that the limiting operator is well-defined; if refinements are taken arbitrarily, in general the limit fails to exist.

Having such a decomposition, we can use it to regularize the classical expression in \eqref{e29}. The regularization enables us to promote \eqref{e29} to an operator through the quantization of the length and the dihedral angle separately. In the following, we define explicitly the two operators $\hat{L}_h^{c}$ and $\hat{\theta}_h^{c}$, combine them to build the curvature operator and study some of its properties in some simple cases.

\subsection{The length operator}\label{sec22}

\begin{figure}[ht]
\centering
\includegraphics[width=0.7\textwidth]{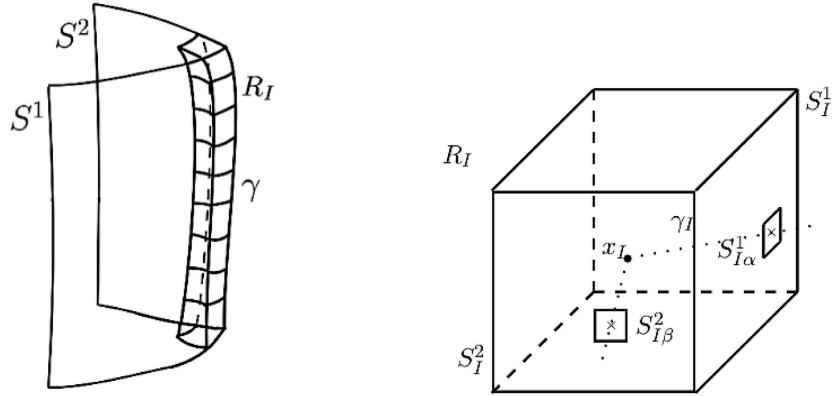}
\caption{Decomposition of the region adjacent to $\gamma$ into cubic cells and the partitioning of the boundary $\partial R_I$ of the cubic cell $R_I$}
\label{fig2}
\end{figure}

In LQG we have three proposals for length operator \cite{lunghezza1, lunghezza2, lunghezza3}. Since our approach to construct a scalar curvature operator is using the dual picture, we choose Bianchi's operator \cite{lunghezza2} which is constructed based on the same dual picture of quantum geometry. Here we summarize the construction contained in \cite{lunghezza2}. The first step in this construction is an external regularization of the classical expression of the length of a curve \eqref{e8}. Considering a region $R$ delimited by two surfaces $S^1$ and $S^2$ intersecting in a curve $\gamma$ parametrized by the variable $s$ (see figure \ref{fig2}), the length of this curve is:
\begin{equation} \label{e11}
L(\gamma)=\lim \limits_{\Delta s \rightarrow 0} \sum \limits_I L_I
\end{equation}
where $L_I$ is the length of a hinge $\gamma_I$ belonging to the boundary of a cell $R_I$ of elementary size $\Delta s$, that is part of a {\it cubic} partition (labeled by $I$) of the region $R$. The length $L_I$ is defined as:
\begin{equation}\label{e30}
\nonumber L_I=\sqrt{\delta_{ij} G_I^i G_I^j}
\end{equation}
with
\begin{equation}\label{e31}
G_I^i=\frac{ \frac{\Delta s}{2} \sum \limits_{\alpha,\beta} \frac{1}{\Delta s^4} Y_{I \alpha \beta}^i}{\sqrt{\frac{1}{8.3!} \sum \limits_{\alpha',\beta', \rho'} \frac{1}{\Delta s^6}\abs{Q_{I {\alpha'} {\beta'} {\rho'}}}}}=\frac{\frac{\Delta s}{2} \sum \limits_{\alpha,\beta} \frac{1}{\Delta s^4}V_{x_I}^{ijk} F_j(S_{I\alpha}^1) F_k(S_{I\beta}^2)}{\sqrt{\frac{1}{8.3!} \sum \limits_{\alpha',\beta', \rho'} \frac{1}{\Delta s^6}\abs{T_{x_I}^{ijk} F_i(S_I^{\alpha'}) F_j(S_I^{\beta'}) F_k(S_I^{\rho'}) }}}
\end{equation}
$\alpha'$, $\beta'$ and $\rho'$ label partitions of $S_I=\partial R_I$, the boundary of the cell $R_I$, while $\alpha$ and $\beta$ label partitions of $S_I^1=R_I \cap S^1$ and $S_I^2=R_I \cap S^2$ respectively (see figure \ref{fig2}). $F_i$ are the fluxes of the electric fields through the surfaces $S^k$:
\begin{equation}
F_i(S^k):=\int \limits_{S^k} \epsilon_{abc}\ E_i^a\ dx^b \wedge dx^c
\end{equation}
$x$ is a coordinate system on $R$. The functions $V_{x_I}^{ijk}$ and $T_{x_I}^{ijk}$ have been incorporated in order to guarantee the $SU(2)$-gauge invariance of the non-local expressions of the nominator and the denominator in \refeq{e31} and they are defined as
\begin{align}
& V_{x_I}^{ijk}=\epsilon^{ij^\prime k^\prime} D^{(1)}(h_{\lambda^1} [A])_{j^\prime} ^{\quad j} D^{(1)}(h_{\lambda^2} [A])_{k^\prime}^{\quad k} \\ 
& T_{x_I}^{ijk}=\epsilon^{i^\prime j^\prime k^\prime} D^{(1)}(h_{\lambda^1} [A])_{i^\prime} ^{\quad i} D^{(1)}(h_{\lambda^2} [A])_{j^\prime} ^{\quad j} D^{(1)}(h_{\lambda^3} [A])_{k^\prime} ^{\quad k} 
\end{align}
where $\lambda^m$ is a curve linking a point $x_I$ inside $R_I$ and a point in $S_{I\alpha}^m$, while $D^{(1)}(h_{\lambda^m})$ is the holonomy of the connection $A$ along the curve $\lambda^m$ taken in the representation ``1'' (the adjoint representation).

Having the regularized expression which has the appropriate classical limit, \eqref{e30} can be promoted to a quantum operator:
\begin{equation}\label{e12}
\widehat{L(\gamma_\omega)}:=\frac{1}{2} \sqrt{\hat{V}_n^{-1}\delta_{ij}\hat{Y}^i(\gamma_\omega)\hat{Y}^j(\gamma_\omega)\hat{V}_n^{-1}}
\end{equation}
The index $\omega=(n,e_1,e_2)$ stands for a wedge (two edges $e_1$ and $e_2$ intersecting in a node $n$) in the graph $\Gamma$ dual to the two faces intersecting in the curve $\gamma$. While $\hat{Y}^i(\gamma_\omega)$ and $\hat{V}_n$ are respectively the two-handed operator and the volume operator:
\begin{align}
& \hat{Y}^i(\gamma_\omega)=\epsilon^{ij^\prime k^\prime} D^{(1)}(h_{e_1} [A])_{j^\prime} ^{\quad j} D^{(1)}(h_{e_2} [A])_{k^\prime}^{\quad k} \hat{F}_j(S_{e_1}) \hat{F}_k(S_{e_2}) \\
& \hat{V}_n=\sqrt{\frac{1}{8.3!} \sum \limits_{\alpha,\beta, \rho} \abs{\epsilon^{i^\prime j^\prime k^\prime} D^{(1)}(h_{e_\alpha} [A])_{i^\prime} ^{\quad i} D^{(1)}(h_{e_\beta} [A])_{j^\prime} ^{\quad j} D^{(1)}(h_{e_\rho} [A])_{k^\prime} ^{\quad k} \hat{F}_i(S_{e_\alpha}) \hat{F}_j(S_{e_\beta}) \hat{F}_k(S_{e_\rho}) }}\label{e13}
\end{align}
note that there exist two versions of the volume operator \cite{RovelliSmolin95, AshtekarLewand98} in the literature; the one presented in \eqref{e13} and used in \cite{lunghezza2} is the Rovelli-Smolin version \cite{RovelliSmolin95} which is consistent with the external regularization scheme. In principle one could use here the version \cite{AshtekarLewand98} even if this is based on a different regularization procedure.

The inverse of the volume operator in \eqref{e12} does not exist a priori but by restricting the domain of the volume operator in \eqref{e13} we get an invertible operator for which we can define an inverse and extended maximally its domain. Considering the geometrical interpretation of such operator, the inverse volume operator must satisfy the following two conditions:
\begin{itemize}
 \item It acts only at the nodes of the spin-network graph and it annihilates spin-network states containing no node;
 \item it has the same eigenstates of the volume operator with non-vanishing eigenvalues equal the inverse of the non-vanishing eigenvalues of \eqref{e13};
\end{itemize}
Such operator exists and can be introduced as
\begin{equation}
 \hat{V}^{-1}:=\widehat{V^{-1}}= \lim \limits_{\epsilon \rightarrow 0} (\hat{V}^2 + \epsilon^2 l^6 )^{-1} \hat{V}
\end{equation}
Where $l$ is a constant which has the dimension of length. This limit is well defined and the result is a hermitian operator $\widehat{V^{-1}}$ which commutes with $\hat{V}$ and admits a self-adjoint extension to the whole Hilbert space.

This length operator measures the length of a curve defined as the intersection of two surfaces dual to two edges sharing a node in a given spin-network graph. This operator is positive semi-definite, hermitian and has a discrete spectrum.

It's important to point out that we can't give a general expression (for any values of the spins) of the length eigenvectors in terms of the intertwiner basis elements because of the presence of the volume operator for which many properties are known but closed formulas are not available \cite{volume} (regardless of the volume operator choice). But of course the eigenstates and eigenvalues can be computed algebraically and numerically for any fixed state.

\subsection{The dihedral angle operator}

We proceed with the same scheme to regularize the expression of the dihedral angle in \eqref{e10} as it was done for the length. We consider a partition that decomposes a region $R$ delimited by two surfaces $S^1$ and $S^2$ intersecting in $\gamma$. Then replacing in \eqref{e10} the contraction of the densitized triad field with the one form normal to the surface $S^k$ by the flux through this surface $F_i(S^k)$ we get the following expression:
\begin{equation} \label{e14}
 \theta_{I\alpha \beta}^{12}= \pi - \arccos \left[ \frac{ \delta^{\prime ik} F_i(S_{I\alpha}^1) F_k(S_{I\beta}^2)}{\sqrt{\delta^{\prime ij} F_i(S_{I\alpha}^1) F_j(S_{I\alpha}^1)} \sqrt{\delta^{\prime kl} F_k(S_{I\beta}^2) F_l(S_{I\beta}^2)}} \right]
\end{equation}
where the functions $\delta^{\prime ik}$ has been inserted in order to guarantee the $SU(2)$-gauge invariance of the non-local expressions of the terms in \refeq{e14} and it's defined as
\begin{equation}
\delta^{\prime ik}=\delta^{i^\prime k^\prime} D^{(1)}(h_{\lambda^1} [A])_{i^\prime} ^{\quad i}\;\; D^{(1)}(h_{\lambda^2} [A])_{k^\prime}^{\quad k}
\end{equation}
$\alpha$ and $\beta$ label partitions of $S_I^1=R_I \cap S^1$ and $S_I^2=R_I \cap S^2$ respectively.

On the quantum level, the fluxes are just the $SU(2)$ generators $\vec{J}$ associated to the edges of the spin-network. Therefore we can write a simple expression for the dihedral angle operator $\hat{\theta}_{ik}$ in the conventional intertwiner basis\footnote{\label{fn} The orthonormal intertwiner states basis $\ket{j_1\dots j_V , \iota_1 \dots \iota_{V-3}}$ for a node of valence $V$,  are basis elements of the Hilbert space $\mathcal{H}_{j_1, \cdots ,j_V}= \text{Inv} [V^{j_1} \otimes \cdots \otimes V^{j_V}]$ with $V^{j}$ being the Hilbert space corresponding to the irreducible representations of $SU(2)$ with spin $j$. Those states are labeled by $V-3$ quantum numbers $\{\iota_i\}$ depending on the coupling of the external legs. We indicate with 
 $\ket{j_{ik}}$ the basis state of a $V$-valent intertwiner with the spins $j_i$ and $j_k$ coupled together and arbitrary couplings for the spins left.}
\begin{align}
 \widehat{\theta(\gamma_\omega)}:=\hat{\theta}_{ik}= \sum \limits_{j_{ik}} \left( \pi - \arccos \left[ \frac{ j_{ik} (j_{ik}+1) - j_i (j_i+1) - j_k (j_k+1)} {2 \sqrt{j_i (j_i+1) j_k (j_k+1)} } \right] \right) \ket{j_{ik}} \bra{j_{ik}}
\end{align}
Where $i$ and $k$ label the two edges forming the wedge $\omega$ dual to the two faces intersecting in the curve $\gamma$. The numbers $j_i, j_k$ and $j_{ik}$ are respectively the values of the spins $i$, $j$ and their coupling.

\subsection{The curvature operator}

Before combining the two operators we defined in the previous sections, let's go back to the dual picture and rebuild our scenario to construct the curvature operator. Considering a spin-network, we build a covering cellular decomposition $\mathscr{C}$ of the 3-space and we focus on a small region which contains only one hinge of the decomposition. The assumption is that the curvature in that region lies on the hinge, this allow us to write the curvature as a combination {\it a la Regge} of the length of this hinge and the deficit angle around it.  This assumption can be seen as a restriction of the phase space of gravity to piecewise flat manifolds, but from our perspective this is rather a different way to regularize our classical expression of the integrated scalar curvature as we also take the continuum limit when the regulator, in this case the typical length, goes to zero. This regularization scheme is well justified classically from the result that any manifold can be arbitrarily approximated by 
a piecewise flat manifold thanks to the convergence of Regge action to the E-H action.

On the other hand, the reason why we chose such regularization scheme is the fact that we can express curvature in terms of simpler quantities that we can handle easily in order to analyze the Hamiltonian constraint and its kernel, which is our final aim.

We have seen that the angle operator is defined without any ambiguities, while for the length operator we still need to think about one issue: the length operator we use is associated only to a node, which means a 3-cell of the covering cellular decomposition, while a hinge can be shared by many 3-cells, which means that for one hinge we have as many length operators as 3-cells sharing it. This is where \eqref{e3} is useful because rewriting the integral over the cellular decomposition as a sum over 3-cells allows us to define our quantities with respect to each 3-cell, and therefore to avoid the ambiguity.

At this point we have to choose an ordering of the operators in the definition of $\widehat{\left[L_h^{c} \left(\frac{2\pi}{\alpha_h}-\theta_h^{c} \right)\right]}$ associated to a hinge $h$ of the 3-cell $c$, and we suggest the following expression:
\begin{align}\label{e15}
\widehat{\left[ L_h^{c} \left(\frac{2\pi}{\alpha_h}-\theta_h^{c} \right) \right]}=\frac{2\pi}{\alpha_h} \hat{L}_h^{c} - \frac{1}{2}( \hat{L}_h^{c} . \hat{\theta}_h^{c}+\hat{\theta}_h^{c} . \hat{L}_h^{c})
\end{align}
\\
Where $\hat{L}_h^{c}=\widehat{L(\gamma_\omega)}$ and $\hat{\theta}_h^{c}=\widehat{\theta(\gamma_\omega)}$ with $\gamma_\omega$ the curve corresponding to the hinge $h$ for the 3-cell $c$ containing the wedge $\omega$. This expression guarantees the operator to be hermitian. Of course this is not the only way to define it, but it's the simplest to think of. Now we can define a quantum curvature operator $\hat{R}_{\mathscr{C}}$ as:
\begin{align}\label{e16}
\hat{R}_{\mathscr{C}}:=\sum \limits_c \sum \limits_{h \in c} \widehat{\left[L_h^{c} \left(\frac{2\pi}{\alpha_h}-\theta_h^{c} \right)\right]} = \sum \limits_c \sum \limits_{h \in c} \frac{2\pi}{\alpha_h} \hat{L}_h^{c} - \frac{1}{2}( \hat{L}_h^{c} . \hat{\theta}_h^{c}+\hat{\theta}_h^{c}. \hat{L}_h^{c})
\end{align}
This operator is the quantum analog \footnote{In the construction of the operator we quantized the classical formula \eqref{e29} restricted to piecewise linear manifolds as an approximation of the continuous ones. We can consider the operator \eqref{e16} also for continuous manifolds; at the quantum level this is not making any difference because the Kinematical Hilbert space of the continuous and the discrete theory are the same (see \cite{Freidel Geiller} for a discussion of how continuous and discrete quantum theories can be seen as a quantization of the same theory in different gauges). However while the operator \eqref{e16} is the quantization of the  exact classical expression representing the integral of the Ricci scalar in the discrete case; the same expression  may fail to be the integral of the Ricci scalar in the continuous case. In this sense one should prove that  the classical formula \eqref{e29} with lengths and angles expressed in terms of fluxes of the Ashtekar Electric field really 
converges to the desired classical expression \cite{wip}.} of the classical expression $\int_{\mathscr{C}} \sqrt{g}R\ dx^3$. It is hermitian and depends on the choice of $\mathscr{C}$. From \eqref{e16}, we see that we can define an operator $\hat{R}_c$ representing the action of $\hat{R}_{\mathscr{C}}$ in the region contained in th 3-cell $c$
\begin{align}\label{e17}
\hat{R}_c:=\sum \limits_{h \in c} \frac{2\pi}{\alpha_h} \hat{L}_h^{c} - \frac{1}{2}( \hat{L}_h^{c} . \hat{\theta}_h^{c}+\hat{\theta}_h^{c}. \hat{L}_h^{c})
\end{align}

Let us now evaluate the action of the operator $\hat{R}_c$ on a cylindrical function $\Psi(\Gamma)$ which is cylindrical with respect to $\Gamma$. Because of the conditions {\it i)} and {\it ii)} on the covering decomposition $\mathscr{C}$, a 3-cell $c$ either contains one node of $\Gamma$ or no node at all. Due to the first condition on the inverse volume operator which is itself used to define the length operator, if the 3-cell $c$ does not contain a node we have
\begin{equation}\label{e18}
 \hat{R}_c \Psi(\Gamma)=0
\end{equation}
If the 3-cell does contain a node, say $n$, then
\begin{align}\label{e19}
\hat{R}_c \Psi(\Gamma) = \sum \limits_{\omega_n^c} \left[ \frac{2\pi}{\alpha_{\omega_n^c}} \hat{L}(\omega_n^c) - \frac{1}{2}( \hat{L}(\omega_n^c) . \hat{\theta}(\omega_n^c)+\hat{\theta}(\omega_n^c). \hat{L}(\omega_n^c))\right] \Psi(\Gamma)
\end{align}
where $\omega_n^c$ labels the wedges\footnote{We recall that a wedge is a set formed by a node and a couple of links sharing it: $\{n,e_i,e_j\}$.} containing the node $n$ and selected by the 3-cell $c$. To make the notation clearer for later, we introduce the coefficient $\kappa(c,\omega_n)$ which is equal to $1$ when the wedge is selected by the 3-cell $c$ and $0$ otherwise. Then \eqref{e19} becomes
\begin{align}\label{e20}
\hat{R}_c \Psi(\Gamma) = \sum \limits_{\omega_n} \kappa(c,\omega_n) \left[ \frac{2\pi}{\alpha_{\omega_n}} \hat{L}(\omega_n) - \frac{1}{2}( \hat{L}(\omega_n) . \hat{\theta}(\omega_n)+\hat{\theta}(\omega_n). \hat{L}(\omega_n))\right] \Psi(\Gamma)
\end{align}
From \eqref{e18} and \eqref{e20} we deduce the action of $\hat{R}_{\mathscr{C}}$ on $\Psi(\Gamma)$
\begin{align}\label{e21}
\hat{R}_{\mathscr{C}} \Psi(\Gamma)=\sum \limits_{n\in \Gamma} \sum \limits_{\omega_n} \kappa(c,\omega_n) \left[ \frac{2\pi}{\alpha_{\omega_n}} \hat{L}(\omega_n) - \frac{1}{2}( \hat{L}(\omega_n) . \hat{\theta}(\omega_n)+\hat{\theta}(\omega_n). \hat{L}(\omega_n))\right] \Psi(\Gamma)
\end{align}
The action of the operator $\hat{R}_{\mathscr{C}}$ depends on the 3-cells containing the nodes of $\Gamma$ (selecting the wedges) and the cells glued to them (fixing the values of the coefficients $\alpha_{\omega_n}$). Hence, it can be kept unchanged as we refine the covering decomposition and shrink the 3-cells to the nodes:
\begin{align}\label{e22}
\lim \limits_{\text{Volume}[c \in \mathscr{C}] \rightarrow 0} \hat{R}_{\mathscr{C}} \Psi(\Gamma)=\sum \limits_{n\in \Gamma} \sum \limits_{\omega_n} \kappa(c,\omega_n) \left[ \frac{2\pi}{\alpha_{\omega_n}} \hat{L}(\omega_n) - \frac{1}{2}( \hat{L}(\omega_n) . \hat{\theta}(\omega_n)+\hat{\theta}(\omega_n). \hat{L}(\omega_n))\right] \Psi(\Gamma)
\end{align}
This limit is well-defined. Unfortunately, this operator \eqref{e22} carries a memory of our choice of the covering decomposition $\mathscr{C}$ through the coefficients $\kappa(c,\omega_n)$ and $\alpha_{\omega_n}$, i.e., the background structure used in the regularization procedure. However one can eliminate the $\kappa(c,\omega_n)$ by {\it averaging} the regularized operator over relevant background structures. While the ambiguity on $\alpha_{\omega_n}$ can be solved by making a suitable choice. We discuss those features in the next section.

\section{Averaging and sharing conditions}\label{sec25}

Following from our construction, the action of the curvature operator on a spin-network depends on the choice of the covering cellular decomposition. This choice is a priori not unique because our prescription to build a decomposition from a spin-network state doesn't imply the uniqueness. We consider this issue as due to the lack of information on a graph enabling the construction of a covering cellular decomposition. Choices of different coverings for a spin-network graph may select different set of the graph wedges, consequently different sharing coefficients $\alpha_h$ and this would imply different results for the action of the curvature operator. 

To remove the first dependence on $\kappa(c,\omega_n)$, we need to appropriately average $\hat{R}_{\mathscr{C}}$ over the relevant background structures, use the resulting operator $\hat{R}_{\mathscr{C}}^{av}$ instead of $\hat{R}_{\mathscr{C}}$ in \eqref{e21} and then take the limit.

\begin{figure}[ht]
\centering
\includegraphics[width=0.9\textwidth]{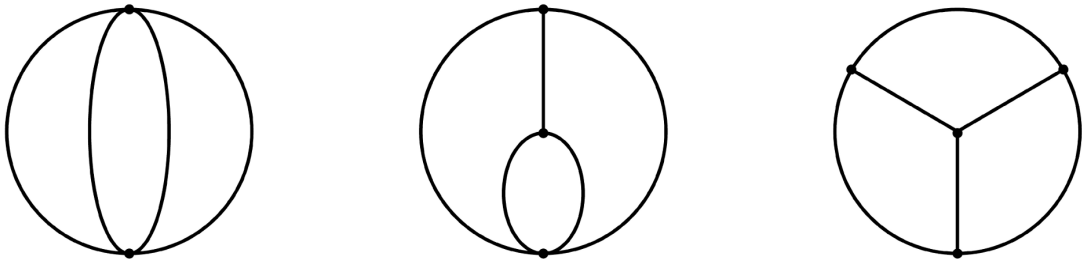}
\caption{\small Classes of spherical polyhedron with 4 faces, from the left to the right: the 4-hosohedron, the class with 3 vertices and the spherical tetrahedron}
\label{class}
\end{figure}

The dependence on $\kappa(c,\omega_n)$ rises directly from the choice of the 3-cells of $\mathscr{C}$ containing the nodes of $\Gamma$. From the definition of the covering decomposition, we can deduce that the 3-cells are isomorphic to spherical polyhedra verifying requirement {\it iv)}. it's important to note that for a fixed number of faces $F$, such spherical polyhedra regroup in a finite number of {\it classes}. A class is defined by the number of edges forming the boundary of each face. For instance, for $F=3$ we have only one class which can be represented by the 3-hosohedron\footnote{A n-hosohedron \cite{PolDiag2} is a tessellation of $n$ areas on a spherical surface such that each area is bounded by two circular arcs and all areas share the same two vertices.}, for $F=4$ we have the 4-hosohedron, the spherical tetrahedron and a third class obtained by taking a 4-hosohedron and replacing one of its vertices by two connected vertices.
These classes can be represented by planar graphs (see figure \ref{class}) similar to Schlegel diagrams for polytopes \cite{PolDiag2}, obtained choosing a face and projecting all the other faces on it as viewed from above.
Labeling the faces $i, j, k \dots$, each class defines, up to permutations of the labels, the adjacency rules for the faces.
Each permutation of the labels of faces defines a configuration. More precisely, if we consider a class and represent each edge in it as $ij$ ($i\neq j$) using the labels of the two faces intersecting at this edge, a configuration is one permutation of faces labels on the full set of edges contained in the class and can be represented as a set of labeled edges $\{ij, kl, mn, \dots \}$. Considering the case $F=4$ as an example, label the faces $1,2,3,4$ and the edges by $ij$, $i\neq j \in\{1,2,3,4\}$. Then the 4-hosohedron (a class) is represented by $\{12,23,34,41\}$ or by any other configuration obtained by permuting the labels of faces. The number of inequivalent configurations for a certain class is of course finite, for the 4-hosohedron is $3$. The tetrahedron is represented by $\{12,13,14,23,24,34\}$ or again by any other configuration obtained by permuting the labels of faces. It's clear that those two classes are defining different configurations hence selecting different edges. 
The fact that the number of inequivalent configurations, we denote it $N_{\text{conf}}$, is always finite allows us to define an averaging procedure over those configurations associated to a 3-cell with a fixed number of faces $F$, or in other words to a node of $\Gamma$ with a given valence $F$.\\

For a given $F$-valent node $n$ of $\Gamma$ , a wedge $\omega$ containing $n$ is considered only in a subset of the full set of configurations, therefore we have a number of appearances $N_{\text{app}}$ of a wedge in the set of configurations, this number depends only on the valence of $n$. Thus we can define a coefficient $\kappa(F_n)$ depending only on the valence $F_n$ of $n$
\begin{equation}
 \kappa(F_n)=\frac{N_\text{app}}{N_\text{conf}} \leq 1
\end{equation}
which stands for the average of $\kappa(c,\omega_n)$. This coefficient is of course the same for all wedges containing the node $n$.

Thus, the action of the averaged operator $\hat{R}_{\mathscr{C}}^{av}$ is
\begin{align}\label{e23}
\hat{R}_{\mathscr{C}}^{av} \Psi(\Gamma)&=\sum \limits_{n\in \Gamma} \kappa(F_n) \sum \limits_{\omega_n} \left[ \frac{2\pi}{\alpha_{\omega_n}} \hat{L}(\omega_n) - \frac{1}{2}( \hat{L}(\omega_n) . \hat{\theta}(\omega_n)+\hat{\theta}(\omega_n). \hat{L}(\omega_n))\right] \Psi(\Gamma)\\ \nonumber
&=\sum \limits_{\omega_n \in \Gamma} \kappa(F_n) \left[ \frac{2\pi}{\alpha_{\omega_n}} \hat{L}(\omega_n) - \frac{1}{2}( \hat{L}(\omega_n) . \hat{\theta}(\omega_n)+\hat{\theta}(\omega_n). \hat{L}(\omega_n))\right] \Psi(\Gamma)
\end{align}
where the sum in the second line is over all the wedges of the graph $\Gamma$.\\

Now we still need to deal with ambiguity on the coefficients $\alpha_{\omega_n}$. Classically the coefficients $\{\alpha_{\omega_n}=\alpha_h\}$ are associated to hinges and they come from the {\it sharing conditions} on the hinges: given a hinge $h$ , we specify all the 3-cells of $\mathscr{C}$ containing this hinge on their boundaries. $\alpha_h$ is the number of those 3-cells. Following from our construction of the operator, those coefficients are totally arbitrary, they depend on the choice of the decomposition but there is no information in the spin-network states that could fix them. They are free parameters. Of course we can always define a prescription to fix them. Nevertheless, the choice of $\{\alpha_{\omega_n}\}$ could control different interesting features of the operator $\hat{R}_{\mathscr{C}}^{av}$, for example the {\ locality} of the operator: we could ask that $\alpha_{\omega_n}=F_n$, that would make the operator ultra-local as it depends only on the properties at the node $n$.
Or we could ask that $\{\alpha_{\omega_n}\}$ are equal to the number of nodes forming the smallest loop in $\Gamma$ containing the wedge ${\omega_n}$ which makes the operator local, and so on. Note that both the AQG framework \cite{17} and the proposal in \cite{Freidel Geiller} for a continuous formulation of the LQG phase space, require the assignment not only of the abstract graphs $\Gamma$ on which the spin-networks are defined, but also a choice of an embedding with a further choice of a dual graph $\Gamma^*$: this assignment in our case would correspond to a unique choice of $\{\alpha_{\omega_n}\}$ that could remove the ambiguity. In \cite{Freidel Geiller} it has been shown that from the discrete data associated with a graph $\Gamma$ it is possible to built a discrete phase space that can be interpreted as the symplectic reduction of the continuous phase space of gravity with respect to a constraint imposing the flatness of the connection everywhere outside of the dual graph $\Gamma^*$. 
This discrete phase space built from $\Gamma$ and $\Gamma^*$ corresponds to {\it''Regge like``} metrics whose curvature is concentrated around not necessarily straight edges. This approach, compatible with the spinfoam perspective, would fix uniquely the covering cellular decomposition and therefore fix the $\{\alpha_{\omega_n}\}$ and remove the need for the averaging procedure described above. However considering the graph $\Gamma^*$ in the quantum theory would require introducing new quantum states that contain information fixing the graph $\Gamma$ and its dual $\Gamma^*$, information which is not given by the spin-network states. Therefore we don't assume in the construction of the operator any particular knowledge of the dual graph $\Gamma^*$.

For now, we say that we choose a prescription fixing the $\{\alpha_{\omega_n}\}$, therefore we can express the action of the final {\it curvature operator} $\hat{R}$ which does not depend on the decomposition as
\begin{align}\label{e24}
\hat{R}\Psi(\Gamma)=\sum \limits_{\omega_n \in \Gamma} \kappa(F_n) \left[ \frac{2\pi}{\alpha_{\omega_n}} \hat{L}(\omega_n) - \frac{1}{2}( \hat{L}(\omega_n) . \hat{\theta}(\omega_n)+\hat{\theta}(\omega_n). \hat{L}(\omega_n))\right] \Psi(\Gamma)
\end{align}

In the following section we present some properties of this operator and discuss its semi-classical limit on some simple cases.

\section{Properties of the curvature operator}

\subsection{Gauge and diffeomorphism transformations}

The curvature operator $\hat{R}$ is $SU(2)$ gauge invariant as a result of introducing the functions $V_{x_I}^{ijk}$, $T_{x_I}^{ijk}$ and $\delta^{\prime ik}$ in the regularized expressions of the length \refeq{e31} and the dihedral angle \refeq{e14}. Therefore, the operator is naturally defined in the space of gauge invariant cylindrical functions. 
Also, since the classical integral over $\Sigma$, considered as an observable, is $\Sigma$-diffeomorphism invariant, we require the same for our operator and this is satisfied by the construction we introduced, thanks to the averaging procedure. Thus it defines the operator in the space of diffeomorphism invariant states. Also, we can think of restricting the domain of integration to an open region $\mathcal{B}$ of the space $\Sigma$, this induces an operator $\hat{R}_\mathcal{B}$ 
\begin{align}\label{e25}
\hat{R}_\mathcal{B} \Psi(\Gamma)=\sum \limits_{\omega_n \in \Gamma \cap \mathcal{B}} \kappa(F_n) \left[ \frac{2\pi}{\alpha_{\omega_n}} \hat{L}(\omega_n) - \frac{1}{2}( \hat{L}(\omega_n) . \hat{\theta}(\omega_n)+\hat{\theta}(\omega_n). \hat{L}(\omega_n))\right] \Psi(\Gamma)
\end{align}
which is still gauge invariant but it is diffeomorphism covariant instead of being invariant. This is simply due to the fact that the action of an arbitrary diffeomorphism does not preserve the region $\mathcal{B}$.

\subsection{Spectrum of the curvature operator}

\begin{figure}[ht]
\centering
\includegraphics[width=0.9\textwidth]{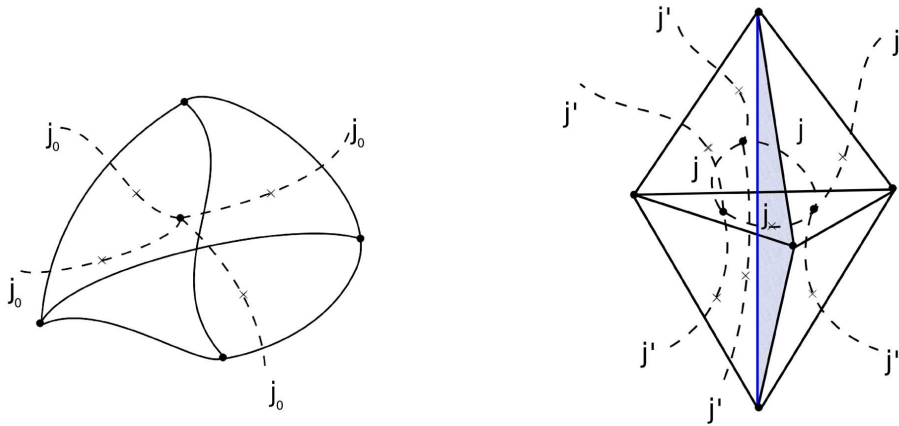}
\caption{\small Examples of tested configurations: 4-valent node with equal spins $j_0$ and three connected 4-valent nodes with equal internal spins $j$ and equal external spins $j'$.}
\label{fig4}
\end{figure}

\begin{figure}[ht]
\centering
\includegraphics[width=1\textwidth]{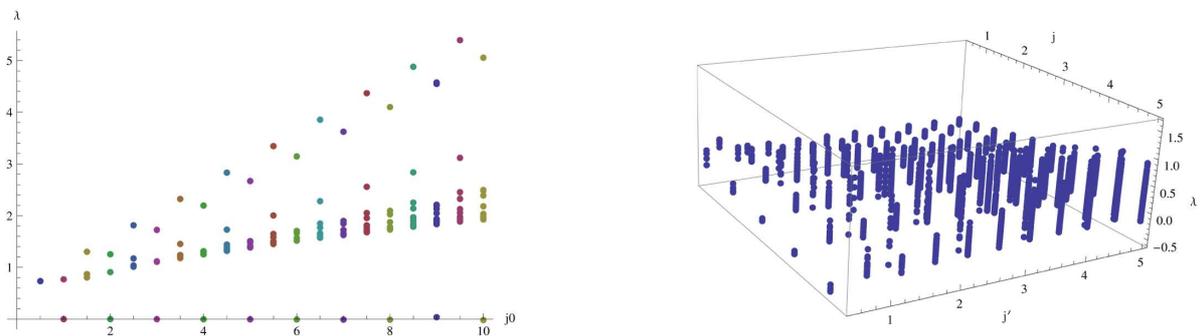}
\caption{\small Spectrum $\lambda$ of the curvature operator: on the left the case of a regular four-valent node plotted as a function of the spin $j_0$. The parameter $\alpha$ is fixed to $1$. On the right the case of the internal geometry in a configuration of three four-valent nodes plotted as a function of the spin $j$ (internal spin) and $j'$ (external spin). The parameter $\alpha$ is fixed to $3$. Units $(8 \Pi \gamma L_P^2)^\frac{1}{2}$ are used.}
 \label{fig5}
\end{figure}

In figure \ref{fig5} we report the eigenvalues of the curvature operator in the case of a four-valent node with all spins equal $j_1 = j_2 = j_3 = j_4 = j_0$, and for the geometry dual to a loop of three four-valent nodes with equal internal spins (labeling the links forming the loop) and equal external spins (see figure \ref{fig4}). This last configuration is an example of a specific choice of a dual graph $\Gamma^*$ which selects a covering decomposition consisting of three glued 4-faces cells fixing $\alpha$ equal to $3$ for the wedges forming the loop.

\subsection{Semi-classical properties}
It is important to stress that in our case the semi-classical limit (large spins limit) does not mean the continuous limit but rather a discrete limit which is classical Regge calculus.

\begin{figure}[ht]
\centering
 \includegraphics[width=0.43\textwidth]{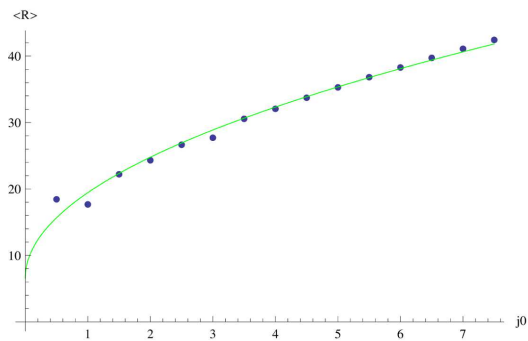}
\caption{\small Expectation values of $\hat{R}$ on Livine-Speziale coherent states plotted as a function of the spin: case of a regular four-valent node ($\alpha=1$). Units $(8 \Pi \gamma L_P^2)^\frac{1}{2}$ are used. The curve is a fit with a square root function $6,57 + 12,87 \sqrt{j_0}$.}
\label{fig6}
\end{figure}

In figure \ref{fig6} we report the expectation values of the curvature operator on Livine-Speziale coherent states \cite{LS} in the case of a regular four-valent node as a function of the spin $j_0$.

Livine-Speziale coherent states are $SU(2)$ invariant intertwiners, obtained by group averaging of SU(2) coherent states \cite{perelemov} which minimize the uncertainty $\Delta=|\langle \vec{J}\ ^2\rangle-\langle \vec{J}\rangle^2|$ in the direction of the angular momentum. The SU(2) coherent states are constructed from the highest weight state through the group action and they are labeled by the spin $j$ and a unit vector $\hat{n}$ defining a direction on the sphere $S^2$. A Livine-Speziale coherent state can be decomposed in the conventional basis of intertwiners\footnotemark[5] as
\be
\ket{j,\hat{n}}_0=\sum \limits_{m_1 \dots m_V} \prod \limits_{i=1}^V a_{m_i}(\hat{n}_i) \sum \limits_{\iota_1 \dots \iota_{V-3}} C_{m_1 \dots m_V}^{\iota_1 \dots \iota_{V-3}}\ \ket{j_1\dots j_V , \iota_1 \dots \iota_{V-3}}
\ee
 where $C_{m_1 \dots m_V}^{\iota_1 \dots \iota_{V-3}}$ are the (generalized) Clebsch-Gordan coefficients and $a_{m_i}(\hat{n}_i)$ are the coefficients defining a coherent state associated to one spin $j_i$ in terms of the spin basis. Using recoupling theory, these generalized coefficients \cite{brink} can always be decomposed into sums of products of conventional (3-valent) Clebsch-Gordan coefficients.
 \\

\begin{figure}[ht]
\centering
 \includegraphics[width=1\textwidth]{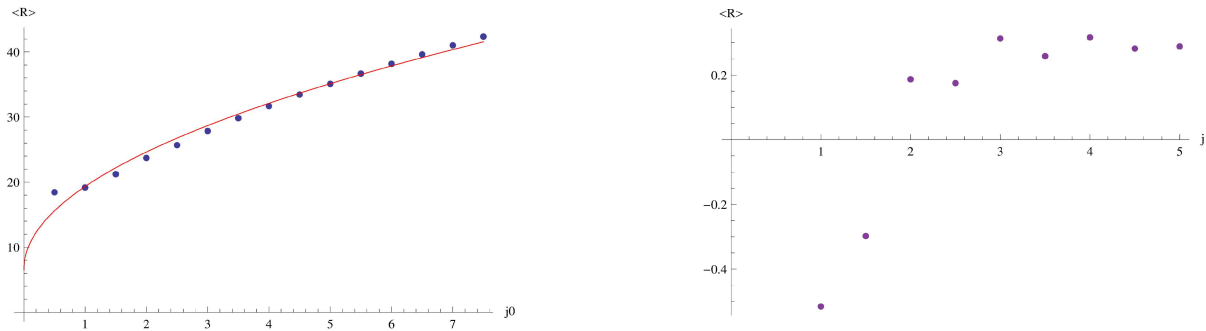}
\caption{\small Expectation values of $\hat{R}$ on Rovelli-Speziale semi-classical states plotted as a function of the spin for two different configurations: on the left the case of a regular four-valent node ($\alpha=1$). The curve is a fit with a square root function $6,55 + 12,77 \sqrt{j_0}$. On the right the case of three four-valent nodes (three tetrahedra glued together) with equal internal spins and equal external spins ($\alpha=3$). Units $(8 \Pi \gamma L_P^2)^\frac{1}{2}$ are used.}
\label{fig7}
\end{figure}

In Figure \ref{fig7} we report the expectation values of the curvature operator on Rovelli-Speziale \cite{CS} semi-classical tetrahedra as a function of the spin in the case of a regular four-valent node (figure on the left in \ref{fig7}), and for the internal geometry in the case of three four-valent nodes with equal internal spins and equal external spins (figure on the right in \ref{fig7}).

Rovelli-Speziale semi-classical tetrahedron is a semiclassical quantum state corresponding to the classical geometry of the tetrahedron  determined by the areas $A_1, \dots, A_4$ of its faces and two dihedral angles $\theta_{12}, \theta_{34}$  between $A_1$ and $A_2$ respectively $A_3$ and $A_4$. It is defined as a state in the intertwiner basis\footnotemark[5] $\ket{j_{12}}$
\be
\psi = \sum \limits_{j_{12}} c_{j_{12}} \ket{j_{12}}
\ee
with coefficients $c_{j_{12}}$ such that
\be
<\hat{\theta}_{ij}> \rightarrow \theta_{ij} \qquad ; \qquad \frac{<\Delta \hat{\theta}_{ij}>}{<\hat{\theta}_{ij}>} \rightarrow 0
\ee
in the large scale limit, for all $ij$. The large scale limit considered here is taken when all spins are large.
\\
The expression of the coefficients $c_{j_{12}}$ satisfying the requirements is:
\be
c_{j_{12}}(j_0,k_0)=\frac{1}{(2\pi \sigma_{j_{12}})^\frac{1}{4}} \exp \left\{ -\frac{(j_{12}-j_0)^2}{4\sigma_{j_{12}}} + i \phi (j_0,k_0) j_{12} \right\}
\ee
where $j_0$ and $k_0$ are given real numbers respectively linked to $\theta_{12}$ and $\theta_{34}$ through the following equations:
\be
j_0^2=2 j_1 j_2 \cos \theta_{12} + j_1^2+ j_2^2 \qquad ; \qquad k_0^2=2 j_3 j_4 \cos \theta_{34} + j_3^2+ j_4^2
\ee
$\sigma_{j_{12}}$ is the variance which is appropriately fixed and the phase $\phi(j_0,k_0)$ is the dihedral angle to $j_0$ in an auxiliary tetrahedron related to the asymptotic of the $6j$ symbol performing the change of coupling in the intertwiner basis (see \cite{CS}).

For a classical regular tetrahedron, using the expression \eqref{e1} for Regge action, the integrated classical curvature scales linearly in terms of the length of its hinges because the angles do not change in the equilateral configuration when the length is rescaled, which means that the integrated classical curvature scales as square root function of the area of a face.  In figures \ref{fig6} and \ref{fig7}(left plot), we see that the expected values of $\hat{R}$ on coherent states and semi-classical (regular) tetrahedra for large spins scales as a square root function of the spin, this matches nicely the semi-classical evolution we expect. In the second case, represented in \ref{fig7}(right plot), in which the state is picked on the configuration where three identical tetrahedra are glued together in flat space to form two glued tetrahedra as in the 2-3 Pachner move (see figure \ref{fig4}), we can notice that the expectation values approach zero as the value of the spins increase which means that the 
configuration in the considered region is close to the classical flat geometry and that is exactly the  expected semi-classical behavior.

\section{Discussion and Outlooks}

In this paper we presented the construction of a curvature operator $\hat{R}$ associated with an open region of a 3-manifold, based on an ``external'' regularization scheme using Regge calculus. We discussed some of its properties and checked its semi-classical behavior in some simple cases. The regularization scheme we adopt in this construction is quite different from the one used in the construction of the volume operator for instance, because  in our case the classical expression is written as the limit of a Regge like discretization instead of introducing a Riemannian sum. Once the regularization is done we express the lengths and angles appearing in the Regge formulas in terms of the elementary variables of the theory, i.e the two-dimensionally smeared triads, which have unambiguous quantum analogs. Thereby we promote the regulated classical expressions to quantum operators and finally remove the regulator.
We choose an appropriate ordering of the length and angle operators to make the resulting curvature operator Hermitian. The quantum operator we find still carries a memory of the background structures used in the regularization procedure. This additional structure can be removed with an averaging procedure over the relevant regularization structures. This construction leads to a well defined, non graph changing, operator $\hat{R}$ up to the choice of some coefficients $\{\alpha_{\omega_n}\}$ specifying, given a spin-network graph, the adjacency relations of the covering cellular decomposition.

As discussed in section \ref{sec25}, the freedom in the choice of the coefficients $\{\alpha_{\omega_n}\}$ is due to the non uniqueness of the covering and more precisely to the non uniqueness of the dual to the spin-network graph. On one hand this tells us that these coefficients can be used to control the properties of the locality of the operator. In this sense the information given by a one cell containing a node is not enough to define unambiguously the curvature operator in the region containing only that node. This picture reminds us of the parallel transport on an infinitesimal closed loop as a way to probe curvature classically in a point on a manifold: the loop is a non local object allowing to explore a very small neighborhood of the relevant point. In the same way, we need to explore the structure around each non empty cell to know the coefficients $\{\alpha_{\omega_n}\}$.
On the other hand, the non uniqueness of a covering cellular decomposition can be fixed by introducing any consistent prescription and a priori the only criterion available to favor a choice over another is the semi-classical limit, but it appears that at least the global behavior of the large spin limit is not affected by this choice.

The regularization scheme we developed exposes a picture in which the geometry is understood as being locally flat and this feature may suggest that this operator is basically defined for the specific class of piecewise flat manifolds. However the fact that classically the limit of Regge expression for the integrated curvature can be taken in such a way that it converges to the integral of the continuous scalar curvature on a chosen manifold is the argument supporting our perspective in which this operator, as implemented in LQG context, is an operator which measures curvature.

This work is mainly motivated by our desire to develop a different way to implement the Hamiltonian operator for the Lorentzian case in LQG. Our hope is that the curvature operator we built will allow us to construct a more analytically manageable Hamiltonian operator and will give the possibility to construct solutions in order to get more insights on the dynamical sector of LQG both in the full theory \cite{Domagala:2010bm, Husain:2011tk} and in simplified models \cite{io1,io2,io3,io4}. This will be the subject of upcoming works.

\section{Acknowledgements}
The authors would like to thank Andrea Dapor and Norbert Bodendorfer for useful comments and interesting discussions. The work  was supported by the grant of Polish Narodowe Centrum Nauki nr 2011/02/A/ST2/00300.

\end{document}